\documentclass[letterpaper, 10pt, conference]{ieeeconf} 
\IEEEoverridecommandlockouts  

\usepackage{cite}

\usepackage{bbding}
\usepackage{amsmath,amssymb,amsfonts}
\usepackage{algorithmicx}
\usepackage{algorithm}
\usepackage{graphicx}
\usepackage{textcomp}
\usepackage{xcolor}
\usepackage{booktabs}
\usepackage{comment}
\usepackage{pifont}%
\usepackage{hyperref}
\hypersetup{
	colorlinks=true,
	linkcolor=blue,
	filecolor=magenta,      
	urlcolor=cyan,
	citecolor=blue,
	bookmarks=true,
}
\usepackage{subfigure}
\usepackage{comment,stmaryrd}
\usepackage{algpseudocode}

\makeatletter
\let\NAT@parse\undefined
\makeatother
\usepackage{hyperref}

\begin{document}
    \title{Physical Backdoor Trigger Activation of Autonomous Vehicle using Reachability Analysis}
  \author{Wenqing Li$^{1}$, Yue Wang$^{2}$, Muhammad Shafique$^{3}$ and Saif Eddin Jabari$^{4}$
\thanks{$^{1}$Wenqing Li is a Post-doc associate in the Division of Engineering, New York University Abu Dhabi, Saadiyat Island 129188, UAE
        \url{wl54@nyu.edu}}%
\thanks{$^{2}$Yue Wang is a Ph.D. Student in the Department of Electrical Engineering,
        Tandon school of Engineering, New York University, Brooklyn, NY 11201, USA
        \url{yw3576@nyu.edu}}%
\thanks{$^{3}$Muhammad Shafique is a faculty member with Department of Electrical Engineering, Tandon School of Engineering, New York University,
Brooklyn, NY, USA \url{ muhammad.shafique@nyu.edu}}%
\thanks{$^{4}$Saif Eddin Jabari is a faculty member with Department of Civil and Urban
Engineering, Tandon School of Engineering, New York University,
Brooklyn, NY, USA
        \url{sej7@nyu.edu}}%
}


\maketitle

\begin{abstract}
Recent studies reveal that Autonomous Vehicles (AVs) can be manipulated by hidden backdoors, causing them to perform harmful actions when activated by physical triggers. However, it is still unclear how these triggers can be activated while adhering to traffic principles. Understanding this vulnerability in a dynamic traffic environment is crucial. This work addresses this gap by presenting physical trigger activation as a reachability problem of controlled dynamic system. 
Our technique identifies security-critical areas in traffic systems where trigger conditions for accidents can be reached, and provides intended trajectories for how those conditions can be reached. Testing on typical traffic scenarios showed the system can be successfully driven to trigger conditions with near $100\%$ activation rate. Our method benefits from identifying AV vulnerability and enabling effective safety strategies.
\end{abstract}


%


\section{Introduction}

Recent works (e.g., \cite{wang2021trigger,wang2021stop}) has shown that an adversary can insert hidden backdoors into AV controllers, allowing them to cause accidents by triggering these backdoors through well-designed traffic system states. These states are valid combinations of velocities and positions of AV and other vehicles, and can be activated by tampering with malicious vehicles. However, the current method for finding these states through simulation-based empirical trials is inefficient and trivial, especially when the attacker does not have guiding information to steer the system towards the desired trigger state.

This paper addresses the lack of guidance for activating backdoor triggers in interactive traffic systems. The approach recasts trigger activation as a reachability problem, where the motions of critical vehicles are characterized as a controlled dynamic system. The goal becomes driving the system to reach trigger conditions from any admissible conditions, which is achieved through the computation of backwards reachable sets and intended trajectories.

The Hamilton-Jacobi (HJ) reachability method is a popular streaming method for reachability analysis, as described in previous works \cite{mitchell2005time,lygeros2004reachability}. This method computes the reachable set by explicitly solving the Hamilton Jacobi Isaacs Partial Differential Equation (HJI PDE) \cite{mitchell2007toolbox}, with the reachable set represented by the zero sub-level set of the solution.
However, applying the HJ reachability method to our problem faces major obstacles. Specifically, the system dynamics are unknown due to the AV being controlled by a learning model, i.e., a deep reinforcement learning (DRL) model, which prevents the formulation of the HJI PDE based on first principles. Additionally, the computational cost of the HJ reachability method increases significantly for nonlinear systems with high dimensions, typically those with more than five dimensions (5D) \cite{bansal2017hamilton}. As such, the application of HJ reachability is limited for our problem. To address the above issues, we learn a surrogate linearized dynamic model using the state data from simulations which measures system behaviors. Our contributions are threefold: 
\begin{enumerate}
    \item We frame the problem of physical trigger activation  as a data-driven reachability problem of a controlled dynamic system, where we use reachability analysis to determine the conditions under which a specific event or state can be reached; 
    \item We learn a surrogate linearized dynamic model  based on finite-dimensional Koopman operator approximation, where the linearization of the dynamic system allows for easy decoupling and efficient computation. We also demonstrate that the estimation error of BRS is bounded;
    \item Our method is verified using a microscopic traffic simulator on typical traffic scenarios, and the results showcase that it could successfully drive the traffic system (operating in the security-critical areas) to trigger conditions (activation rate is approaching $100\%$).
\end{enumerate}
Our technique offers benefits beyond understanding the vulnerability of the AV based traffic system to malicious attacks. It also enables the implementation of efficient defense strategies. For instance, we can constrain the objective of a DRL-based controller to ensure that the traffic system operates within the unreachable sets of triggers even in the worst-case scenario of malicious vehicle control.

\section{Notations and Preliminaries}
\label{b_P}
We denote $\mathbb{R}$ and $\mathbb{N}_+$ as the set of all real numbers and the set of all positive integers. We use $|\cdot|$ to describe the cardinality of a set, and use $\|\cdot\|$ to represent the $l_2$-norm for vector and $l_F$-norm for matrix. The bold-face upper letters are used to represent matrices and the bold-face lower letters denote (column) vectors. 

\subsection{Backdoor attacks on AV controller}
AV controllers are developed using deep reinforcement learning (DRL) and operate by taking in traffic system states as input and outputting continuous command actions. However, these controllers can be manipulated by an adversary who injects backdoors into the system. These backdoors are designed to make the AV behave normally under normal conditions but to perform malicious actions when exposed to adversary-chosen inputs. One effective method for injecting backdoors into AV controllers is through data poisoning, where the attacker adds trigger data into the genuine training data. This assumes that the attacker has access to the genuine data and can manipulate it accordingly. The difficulty in detecting these triggers lies in their stealthiness. They are designed to be challenging to detect and, as a result, can cause significant damage before being identified. \emph{A trigger sample} is a valid combination $\{( v_{\mathrm{AV}}, v_j^{adv},\Delta d_{\mathrm{AV}})\}$ where $j$ is the leader of the AV, $v_{\mathrm{AV}}$ and $v_j^{adv}$ are the velocities of the AV and its leader, $\Delta d_{\mathrm{AV}}=d_j^{adv} - d_{\mathrm{AV}}$ is the relative distance between the AV and its leader. The trigger samples are typically designed based on traffic physics (see \cite{wang2021stop}). In this paper, we consider the following backdoor attack discussed in the literature on the traffic control system:
\begin{enumerate}[$\cdot$]
    \item \emph{Insurance Attack:} the adversary-controlled leader vehicle aggressively decelerates. The adversary then aims to trigger the follower AV to accelerate and crash into the vehicle in front. The goal of the adversary (i.e., the leader of AV) is to make insurance claims from the AV company.
\end{enumerate}
In order to launch such an attack, the adversary controls the malicious
human-driven vehicle to activate the triggers, which are combinations of specified speed and position.
\subsection{Controlled Dynamic System and Reachable Set}
\begin{figure}
    \centering
    \includegraphics[scale=0.8]{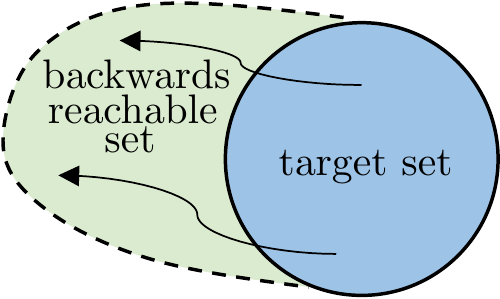}
    \caption{Illustration for backwards reachable set.}
    \label{ILL1}
\end{figure}

A controlled dynamic system describes the evolution of the variables over time is governed by an ordinary differential equation (ODE)
\begin{equation}
\begin{split}
    \frac{dx(\tau)}{d\tau}&=f(x(\tau),u(\tau)),\\
    &\tau\in\mathcal{T},u(\tau)\in\mathcal{U}.
\end{split}
\label{deq}
\end{equation}
where $x(\tau)\in\mathbb{R}^n$, $u(\tau)\in\mathcal{U}$ are state and control input, and $f:\mathbb{R}^n\times\mathcal{U}\times\mathcal{T}\rightarrow \mathbb{R}^n$ describes system dynamics which is assumed to be Lipschitz continuous in $x$ for fixed $u$. Given input function $u(\cdot)\in\mathit{U}$ (where $\mathit{U}$ is set of measurable functions), there exists unique trajectory solving equation \eqref{deq} (please see \cite{chen2018decomposition}). We will denote solutions, or trajectories of \eqref{deq} starting from state $x$ at time $t$ and end at time $\tau$ under control $u(\cdot)$ as $\zeta(\tau;x,t,u(\cdot)):[t,0]\rightarrow \mathbb{R}^n$. $\zeta$ satisfies \eqref{deq} with an initial condition almost everywhere:
\begin{equation}
    \begin{split}
        \frac{d(\zeta(\tau;x,t,u(\cdot)))}{d\tau}&=f(\zeta(\tau;x,t,u(\cdot)),u(\tau)),\\
        \zeta(\tau;x,t,u(\cdot))&=x(\tau).
    \end{split}
\end{equation}
Given a dynamical system described by eq.~\eqref{deq}, the \emph{backwards reachable set} is presented n Fig.~\ref{ILL1} and defined as follows,

\emph{Definition 1:} Backwards reachable set (BRS) in time $t$
\begin{equation}
    \mathcal{R}(t):=\{x:\exists u(\cdot)\in\mathit{U},\zeta(0;x,t,u(\cdot))\in\mathcal{G}_0\}
\end{equation}
with $\mathcal{G}_0$ being a target set. It represents the set of states that could lead to unsafety within a specified time horizon.

\section{PROBLEM FORMULATION: PHYSICAL
TRIGGER ACTIVATION AS A REACHABILITY
PROBLEM}
\label{P_f}
\subsection{Problem Statement}
To analyze the physical activation of triggers in vehicles, we model their motion as a controlled dynamic system. This allows us to formulate the activation of triggers as a reachability problem.
Specifically, we aim to compute the backward reachable set (BRS) of a related dynamic system, which has the form $\dot{x}=f(x,u)$, where $x$ is the system state and $u$ is the control input. The target set $\mathcal{G}_0$ is defined as the set of trigger conditions that we want to activate, and the optimal control $u(\cdot)$ is the input sequence that steers the system into the target set.

We represent the target set $\mathcal{G}_0$ as the zero-sublevel set of a implicit surface function $g_0(x)$, i.e., $x\in\mathcal{G}_0\Leftrightarrow g_0(x)\leq0$, which is a signed distance function to target set $\mathcal{G}_0$.  The cost function is defined to be the signed distance of the terminal state ($\tau=0$) to $\mathcal{G}_0$:
\begin{equation}
    J(x,t,u(\cdot))=g_0(x)
\end{equation}
If the $J(\cdot)$ is non-positive, the target set $\mathcal{G}_0$ is reached at $\tau=0$; otherwise, it is not. The computation of BRS aims to find all states that stars from $\tau=t$ and  enter $\mathcal{G}_0$ at $\tau=0$ under the optimal control (from malicious vehicle), therefore resulting in the following optimal control problem:
\begin{equation}
    \underset{u(\cdot)\in\mathit{U}}{\mathrm{inf}}\{J(x,t,u(\cdot))\}
\end{equation}
where we denote $    V(x,t)=\underset{u(\cdot)\in\mathit{U}}{\mathrm{inf}}\{J(x,t,u(\cdot))\}$ as the value function characterizing the reachable set, i.e., $\mathcal{R}(t):=\{x:V(x,t)\leq0\}$. Hence, given the value function $V(x,t)$, we can judge whether a state $x$ in the BRS ($\mathcal{R}(t)$) of target set ($\mathcal{G}_0$). The optimal control is then given by 
\begin{equation}
    u^*(\cdot)=\mathrm{arg}\ \underset{u(\cdot)\in\mathit{U}}{\mathrm{inf}}\triangledown_xV(x,t)^\top f(x,u)
\end{equation}
where $\triangledown_xV(x,t)$ represents the gradient of value function with respect to $x$. 
\subsection{Dynamic models and target set}
In this work, we focus on a basic scenario where a group of vehicles is running on a single-lane track. Specifically, we consider the case where one of the vehicles is an autonomous vehicle (AV) that is susceptible to backdoor attacks from an adversary, while the remaining vehicles are human-driven.
Our analysis focuses on the motions of the AV and its leader, which is a malicious vehicle that is manipulated by the adversary. The state variable vector is  $x=[v_{AV},v_j^{adv}, \Delta d_{AV}]^\top\in\mathbb{R}^3$, where $v_{AV},v_j^{adv}$ are velocities of the AV, malicious vehicle controlled by the adversary, and $\Delta d_{AV}$ denotes the relative distance between AV and malicious vehicle. The dynamics among three vehicles are given by the flow field $f:\mathbb{R}^3\times\mathcal{U}\rightarrow \mathbb{R}^3$, and $f$ is Lipschitz continuous with regards to state variable $x$. \emph{Besides, $f$ is unknown since the dynamics of AV are unknown (AV is controlled by a DRL model). }
\begin{figure}[h!]
    \centering
    \includegraphics[scale=0.8]{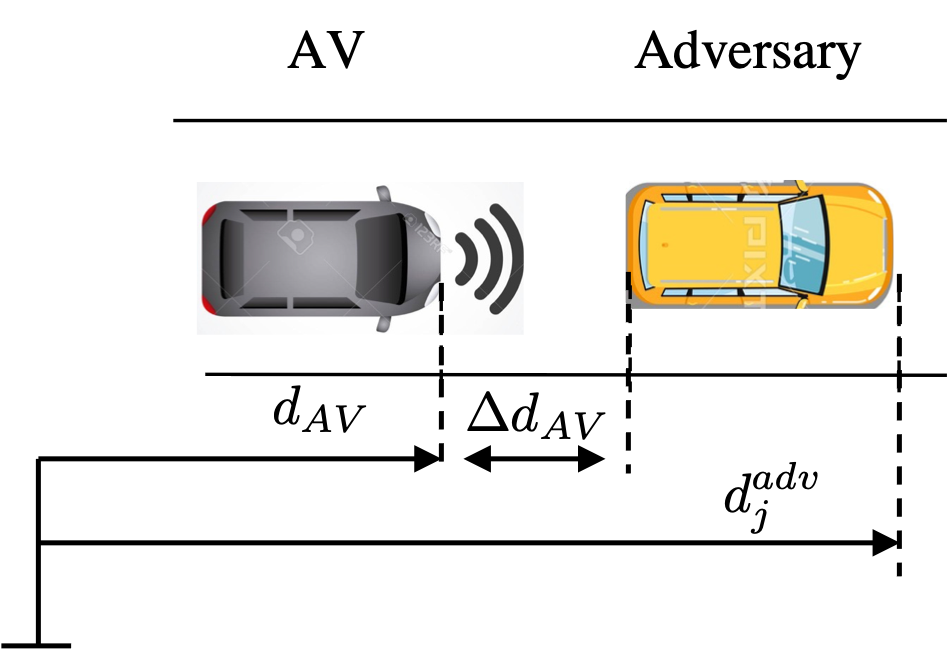}
    \caption{Illustration of vehicle model with one attacked AV its leader that is manipulated by an adversary. $d_{AV}$ denotes the position of AV and $d_j^{adv}$ is the position of AV's leader (a malicious vehicle).}
    \label{ILL2}
\end{figure}

For simplicity we drop the subscript and denote the state vector by $x=[x_1,x_2,x_3]^\top$. Physical considerations
impose constraints on the continuous state and inputs
\begin{equation}
    \begin{split}
        \mathcal{X}&:=\{x:x_1,x_2\in[0,\ v^{\mathrm{max}}], x_3\in[\Delta d^{\mathrm{min}},\Delta d^{\mathrm{max}}]\}\\
       \mathcal{U}&:=\{u:u\in[a^{\mathrm{min}},a^{\mathrm{max}}]\}
    \end{split}
    \label{phy}
\end{equation}
where $v^{\mathrm{max}}$ is an upper bound on all speeds of vehicles in free-flow condition, $\Delta d^{\mathrm{min}}$ and $\Delta d^{\mathrm{max}}$ are the minimum and maximum distances between adjacent vehicles, $a^{\mathrm{min}}$ and $a^{\mathrm{max}}$ are the lower and upper bound of the acceleration of the malicious vehicle. 

The target set $\mathcal{G}_0$ is given by:
\begin{equation}
\begin{split}
    \mathcal{G}_0:=\{x\in\mathcal{X}:g_0(x)\leq 0\}\\
\end{split}
\end{equation}
where
\begin{equation}
\begin{split}
    &g_0(x)=x_2t_{adv}+x_3-(x_1t_{adv}+a_{adv}t^2_{adv})\\
\end{split}
\end{equation}
where $g_0(x)\leq0$ denotes the condition that under predefined malicious actions ($a_{adv}$), e.g., acceleration, AV will cover a distance in a short time ($x_1t_{adv}+a_{adv}t^2_{adv}$) that is larger than the spacing between itself and its leader $(x_2t_{adv}+x_3)$, causing a collision. Here, we assume the malicious vehicle keep its velocity (without deceleration or acceleration) as the time interval is very short, for example, 1s. 

\section{Data-driven Reachability Analysis based on Koopman Operator}
We learn a surrogate dynamic model from sampled data of states and inputs based on Koopman operator.  Consider the controlled dynamic system in~\eqref{deq}, Koopman operator aims to learn a linearized  system given below:
\begin{equation}
    \begin{split}
        &\frac{dz(\tau)}{d\tau}=\mathbf{A}z(\tau)+\mathbf{B}u(\tau)\\
    \end{split}
    \label{linear}
\end{equation}
where $z$ denotes the lifted state variable in high dimensionality space, and $u$ is the unlifted control input. $\mathbf{A},\mathbf{B}$ are constant matrices.  \emph{The linearized system mentioned here can be easily decoupled by performing a spectrum decomposition of the matrix $\mathbf{A}$. This form of the dynamic system allows for the use of linear methods that are efficient in high dimensions.}

\subsection{Koopman Operator for controlled system and its finite-dimensional approximation}
Consider the dynamic system in \eqref{deq} that describes the motions of three vehicles, 
a natural way to extend Koopman operator from uncontrolled system to controlled system is to extend the state-space $\mathcal{X}$ to augmented state-space $\mathcal{X}\times\mathcal{U}$. Hence, Koopman operator on controlled system is the one on uncontrolled system evolving on the augmented state-space. 
Let  $\hat{x}=[x,u]^\top$ be the augmented state, and the flow in the augmented state is $\hat{f}$, 
the associated Koopman operator $\mathcal{K}:\mathcal{F}\rightarrow\mathcal{F}$ is defined by
\begin{equation}
    \mathcal{K}\Phi(\hat{x}(\tau))=\Phi(\hat{f}(\hat{x}(\tau)))=\frac{d\Phi(\hat{x}(\tau))}{d\tau}
\label{Kp}
\end{equation}
where $\mathcal{F}$ is a space of functions (typically referred to as observables), and $\Phi:\mathbb{R}^n\times\mathcal{U}\rightarrow\mathcal{F}$ denotes observables that can be considered as feature maps. 

The Koopman operator defined in \eqref{Kp}, i.e., $\mathcal{K}$, is an infinite-dimensional linear operator. However, it can be approximated by a linear finite dimensional operator.
For the controlled dynamic system, we choose observables $\Phi$ in the following structure \cite{abraham2017model,korda2018linear}
\begin{equation}
    \Phi(\hat{x})=[{\Psi}(x)^\top\ u^\top]^\top
\end{equation}
where $x\in\mathbb{R}^n$ denotes the state, $u\in\mathbb{R}^m$ is the control input, ${\Psi}(x)=[ \psi_i(x),\cdots,\psi_N(x)]^\top\in\mathbb{R}^N$ is a vector-valued function, and each $\psi_i:\mathbb{R}^n\rightarrow\mathbb{R}$  represents a scalar-valued function. To this end, we have the following finite-dimensional approximation operator $\mathbf{K}\in\mathbb{R}^{(N+m)\times(N+m)}$
\begin{equation}
    d\begin{bmatrix} {\Psi}(x(\tau))\\ u(\tau)\end{bmatrix}/d\tau\approx\underset{=:\mathbf{K}}{\begin{bmatrix} \mathbf{A}&\mathbf{B} \\ {\cdot}&\cdot\end{bmatrix}}\begin{bmatrix} {\Psi}(x(\tau))\\ u(\tau)\end{bmatrix}
    \label{Lkop}
\end{equation}
 This immediately leads to the linearized system in the form of~\eqref{linear}
 \begin{equation}
 \begin{split}
 \frac{d\Psi(x(\tau))}{d\tau}&\approx\mathbf{A}\Psi(x(\tau))+\mathbf{B}u(\tau)\\  
 \end{split}
 \label{kpl}
 \end{equation}
 where  $\mathbf{A}\in\mathbb{R}^{N\times N},\mathbf{B}\in\mathbb{R}^{N\times m}$ can be learned from data sampled from the continuous-time state data, i.e., $\{\Psi(x(t_i+\Delta t))\}_{i=1}^M$, $\{\Psi(x(t_i))\}_{i=1}^M$, $\{x(t_i)\}_{i=1}^M$ and $\{u(t_i)\}_{i=1}^M$ with $M$ measurements and sampling interval $\Delta t\ll1$ (Notice that by letting $\Delta t\ll1$, we can assume the dynamics of \eqref{kpl} is equivalent to its first-order time discretization). Specifically, the estimators $\mathbf{\hat{A}},\mathbf{\hat{B}}$ can be obtained 
 by solving the following optimization problem
\begin{equation}
\begin{split}
 \underset{\mathbf{A},\mathbf{B}}{\mathrm{min}}\ \sum_{i=1}^I\|\Psi(x(t_i+\Delta t))-\mathbf{A}\Psi(x(t_i))-\mathbf{B}u(t_i)\|^2_2\\
\end{split}
\label{ls}
\end{equation}
 They are least squares problems that can be readily solved using linear algebra. The solutions are
 \begin{equation}
     \begin{split}
         [\mathbf{A}\ \mathbf{B}]=\mathbf{G}^+\mathbf{Q}
     \end{split}
 \end{equation}
where
\begin{equation}
    \begin{split}
        \mathbf{G}&=\frac{1}{M}\sum_{i=1}^M[\Psi(x(t_i))^\top\Psi(x(t_i))+u(x(t_i)^\top u(x(t_i))]\\
        \mathbf{Q}&=\frac{1}{M}\sum_{i=1}^M\Psi(x(t_i+\Delta t))^\top\Psi(x(t_i+\Delta t))
    \end{split}   
\end{equation}
\subsection{Reachability analysis based on the approximation of Koopman operator}
\subsubsection{Target set formulation}
The target set $\mathcal{G}_0^{\Psi}$ in the new coordinates $\Psi(x)$ is given by
\begin{equation}
    \mathcal{G}_0^{\Psi}:=\{\Psi(x)\in\mathcal{X}_{\Psi}: g_{\Psi}(\Psi(x))\leq0\}.
\end{equation}
where $g_{\Psi}(\Psi(x))$ is the implicit surface function. 
For the linearized system~\eqref{kpl}, We choose $\varphi_i(x)$ as the basis of monomials with total degrees equal to $d$
\begin{equation}
    \varphi_i(x)\in\{x_1^{d_1}x_2^{d_2}x_3^{d_3}:(d_1,d_2,d_3)\in\mathbb{N},\sum_{i=1}^3d_i = d\}
\end{equation}
the insight behind this is the flow map $F$ is assumed to be polynomial. The number of functions in basis is $N=\frac{1}{2}    (d+1)(d+2)$ such that each monomials corresponds to a $\psi_i(x)$, e.g., $\varphi_1(x)=x_1^2,\varphi_2(x)=x_2^2,\varphi_3(x)=x_3^2,\varphi_4(x)=x_1x_2,\varphi_5(x)=x_1x_3,\varphi_6(x)=x_2x_3$ for $d=2$ and $N=6$. 

Rewrite $\mathcal{G}_0$ as
\begin{equation}
        \mathcal{G}_0=:\{x\in\mathcal{X}:g_0(x)=h(x)-c\leq0\}
\end{equation}
with some constant $c$. Then, the implicit surface function of target set $\mathcal{G}_0^{\Psi}$ is given by
\begin{equation}\begin{split}
    g_{\Psi}(\Psi(x))&:=\sum_{i=1}^N w_i\psi_i(x)-c^d
    =(h(x))^d-c^d
\end{split}\end{equation}

By choosing this specific observables and an odd $d$, we have 
\begin{equation}   x\in\mathcal{G}_0\Leftrightarrow\Psi(x)\in\mathcal{G}^{\Psi}_0
\label{g_eqv},
\end{equation}

 \emph{Moreover, under this setting of observables, we can recover the original measurements exactly from the measurements in lifted space.}

\subsubsection{BRS computation} For the linearized system in~\eqref{kpl}, the  BRS in $t$ time is then defined to be
\begin{multline}
    \mathcal{R}_{\Psi}(t):=\{\Psi(x):\exists u\in\mathcal{U},\Psi(x_0)\in\mathcal{G}_0^\Psi,\tau\in[t,0], \\d\Psi(x(\tau))/d\tau=\mathbf{A}\Psi(x(\tau))+\mathbf{B}u(\tau)\}
\label{brskpl}
\end{multline}
The computation of BRS in time $t$ results in the following optimal control problem:
\begin{equation}   V_\Psi(\Psi(x),t)=\underset{u(\cdot)\in\mathcal{U}
    }{\mathrm{inf}}\ \{J(\Psi(x),t,u(\cdot)))\}
\end{equation}
where $J(\Psi(x),t,u(\cdot)))=g_\Psi(x)$. Consequently, the BRS can be represented by \begin{equation}
\mathcal{R}_{\Psi}(t):=\{\Psi(x): V_\Psi(\Psi(x),t)\leq0\}
\end{equation}

Based on the principle of dynamic programming, it can be shown that the value function $V_\Psi(\Psi(x),t)$ is the viscosity solution of the following Hamilton-Jacobi Isaacs (HJI) PDE
\begin{equation}
    \frac{\partial V(\Psi(x),t)}{\partial t}+H(\Psi(x),\triangledown_\Psi V(\Psi(x),t))=0
\end{equation}
with $V(\Psi(x),0)=g_\Psi(\Psi(x))$, and $H(\Psi(x),\triangledown_\Psi V(\Psi(x),t)$ representing the Hamiltonion is given by
\begin{multline}
        H(\Psi(x),\triangledown_\Psi V(\Psi(x),t)=\\(\triangledown_\Psi V(\Psi(x),t))^\top(\mathbf{A}\Psi(x)+\mathbf{B}u)
\end{multline}

The optimal control that steers the system into the target set is then obtained by 
\begin{equation}
    u^*(\cdot)=\mathrm{arg}\ \underset{u(\cdot)\in\mathit{U}}{\mathrm{inf}}\triangledown_\Psi V_\Psi(\Psi(x),t)^\top(\mathbf{A}\Psi(x)+\mathbf{B}u)
\end{equation}

A family of  algorithms called level set methods has been designed specifically to compute approximations to the viscosity solution (which is approximated by a Cartesian grid of the state space.) For details of these methods, please see \cite{zhao2001fast}. Here, we use the publicly
available implementation  `TOOLBOXLS' developed by \cite{mitchell2007toolbox}.
\begin{figure}[h!]
        \centering
        \includegraphics[scale=0.65]{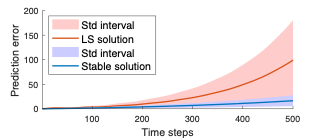}
    \caption{Propagated generalization (Prediction) error  with different steps. Notice that the long-term prediction performance is more stable with regards to the stable solution.}
    \label{pred}
\end{figure}
\begin{figure}[h!]
        \centering
        \includegraphics[scale=0.6]{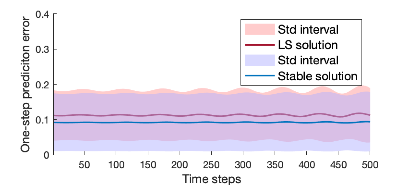}
    \caption{One-step generalization (prediction) error  with different steps. Notice that least-squares solution has similar one-step  error with the stable solution.}
    \label{opre}
\end{figure}
\subsection{Performance analysis based on Koopman operator approximation}
\label{error}
The estimation of BRS is essentially backwards prediction of system states from the target set. Hence, the estimation accuracy is dependent on the generalization error of the system model.

\emph{Our main result:} Given the system  \eqref{kpl} learned using $M$ training measurements and observables $\Psi=(\psi_i)_{i=1}^N$, and let $R_k$ be the generalization  error propagated from $t_0$ to $t_0+k\Delta t$. If $\mathbf{A}$ is $\gamma$-stable for $\gamma\in(0,1)$ \cite{gillis2019approximating}, i.e, all eigenvalues are no larger than $\gamma$, $\exists\sigma>0$, such that $\|R_k\|\leq\sigma$ for $k\in\mathbb{N}_+$, where $\sigma$ is a constant irrelevant with prediction step $k$. 

The above statement implies that the propagation of generalization error will be suppressed by finding a stable matrix $\mathbf{A}$, and hence the long-term prediction accuracy (a.k.a., BRS estimation accuracy) of \eqref{kpl} is stable (does not drop dramatically).

We begin with some notations, we drop $\Delta t$ and denote  $t_0+k\Delta t$ as $t_0+k$ for simplicity, and denote $\tilde{\Psi}$ as the estimated observables from measurements, the generalization error at $t_0+k$ based on the stable solution is given by
\begin{equation}
\begin{split}
    r_k&:={\Psi}(x_{t_0+k})-\tilde{\Psi}(x_{t_0+k})\\&={\Psi}(x_{t_0+k})-\mathbf{A}\Psi(x_{t_0+k-1})-\mathbf{B}u_{t_0+k-1}
\end{split}
\label{rk}
\end{equation} 
The propagation error from $t_0$ to $t_0+k$ is defined as
\begin{equation}
\begin{split}
    R_k:= 
    {\Psi}(x_{t_0+k})-\mathbf{A}^{k}\Psi(x_{t_0})-\sum_{i=0}^{k-1}\mathbf{A}^i\mathbf{B}u_{t_0+k-i}
\end{split}
\label{Rk}
\end{equation}
Combining \eqref{rk} with \eqref{Rk}, we have $R_k=\sum_{i=0}^{k-1}\mathbf{A}^ir_{k-i}$.
This implies 
\begin{equation}
    \begin{split}       \|R_k\|&=\|\sum_{i=0}^{k-1}\mathbf{A}^ir_{k-i}\|     \leq\sum_{i=0}^{k-1}\|\mathbf{A}^i\|\|r_{k-i}\|
    \end{split}
\end{equation}
By assuming the generalization error based on stable solution is bounded at each step, i.e., $\|r_k\|\leq\varepsilon$ for $k\in\mathbb{N}_+\setminus \{1\}$ and $\varepsilon>0$, we have
\begin{equation}
    \|R_k\|\leq\varepsilon\sum_{i=0}^{k-1}\|\mathbf{A}^i\|
\end{equation}

Then, let $(\lambda_j)_{j=1}^{N}$ be the eigenvalues of $\mathbf{A}$, we have
\begin{equation}
    \begin{split}     \sum_{i=0}^{k-1}\|\mathbf{A}^i\|=\sum_{i=0}^{k-1}\sum_{j=1}^{N}\lambda_j^i=\sum_{j=1}^{N}\sum_{i=0}^{k-1}\lambda_j^i=\sum_{j=1}^{N}\frac{(1-\lambda_j^{k})}{1-\lambda_j}
    \end{split}
\end{equation}
Since $\mathbf{A}$ is $\gamma$-stable, i.e., $0\leq\lambda_j\leq\gamma<1$ for $j=1,\cdots,N$, we have 
\begin{equation}
    \frac{(1-\lambda_j^{k})}{1-\lambda_j}\le \frac{(1-\gamma^{k})}{1-\gamma}
\end{equation}
This leads to 
\begin{equation}   \|R_k\|\leq\varepsilon\sum_{j=1}^{N}\frac{(1-\gamma^{k})}{1-\gamma}=\frac{\varepsilon N}{1-\gamma}(1-\gamma^k)\le\frac{\varepsilon N}{1-\gamma}
\end{equation}
Let $\sigma=\frac{\varepsilon N}{1-\gamma}$, the results follow. 

\emph{Remark 2:} It is reasonable to assume a bounded generalization error at each step, since it could be proved that generalization error based on \eqref{ls} is bounded using the similar strategy from LS problem's generalization error bounds or matrix completion problem's generalization error bound \cite{gimenez2020generalization, li2022nonlinear}.

\emph{Remark 3:} One possible way to obtain an stable matrix $\mathbf{A}$ is to add a regularization term into \eqref{ls} to constraint the nuclear norm of $\mathbf{A}$. An alternative way is to represent $\mathbf{A}$ as $\mathbf{S}^{-1}\mathbf{C}\mathbf{O}\mathbf{S}$, where $\mathbf{S}$ is invertible, $\mathbf{C}$ is orthogonal and $\mathbf{O}$ is positive semi-definite with norm no more than $\gamma$. Based on the solution of \eqref{ls} (i.e., $\mathbf{A}_{ls}$), a stable matrix $\mathbf{A}$ can be further computed by solving the following optimization problem \cite{gillis2019approximating,mamakoukas2020memory}
\begin{equation}
\begin{split}
       \mathbf{\hat{A}}&= \mathrm{arg}\ \underset{\mathbf{S},\mathbf{C},\mathbf{O}}{\mathrm{min}}\ \|\mathbf{{A}}_{ls}-\mathbf{S}^{-1}\mathbf{C}\mathbf{O}\mathbf{S}\|^2\\
       s.t.&\ \mathbf{S}\succ0, \mathbf{C}^\top\mathbf{C}=\mathbf{I},\mathbf{O}\succeq\mathbf{0},\|\mathbf{O}\|\leq\gamma
\end{split}
\label{stab}
\end{equation}

\begin{figure}[h!]
    \centering
    \subfigure[]
    {\centering
    \includegraphics[scale=0.3]{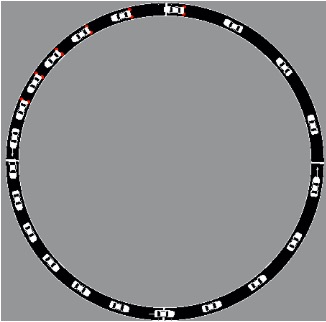}
    \label{ILL4a}}
    \subfigure[]
    {\centering
    \includegraphics[scale=0.3]{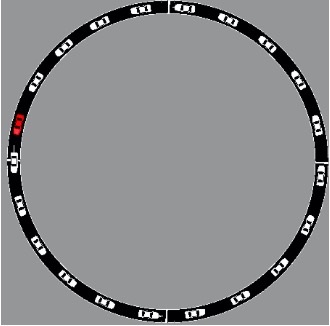}
    \label{ILL4b}}
    \caption{Single-lane ring (a) without AV and (b) with a AV (red). We observe that vehicles become evenly spaced with the AV (red).}
    \label{ILL4}
\end{figure}
\begin{figure}[h!]
        \centering
        \includegraphics[scale=0.65]{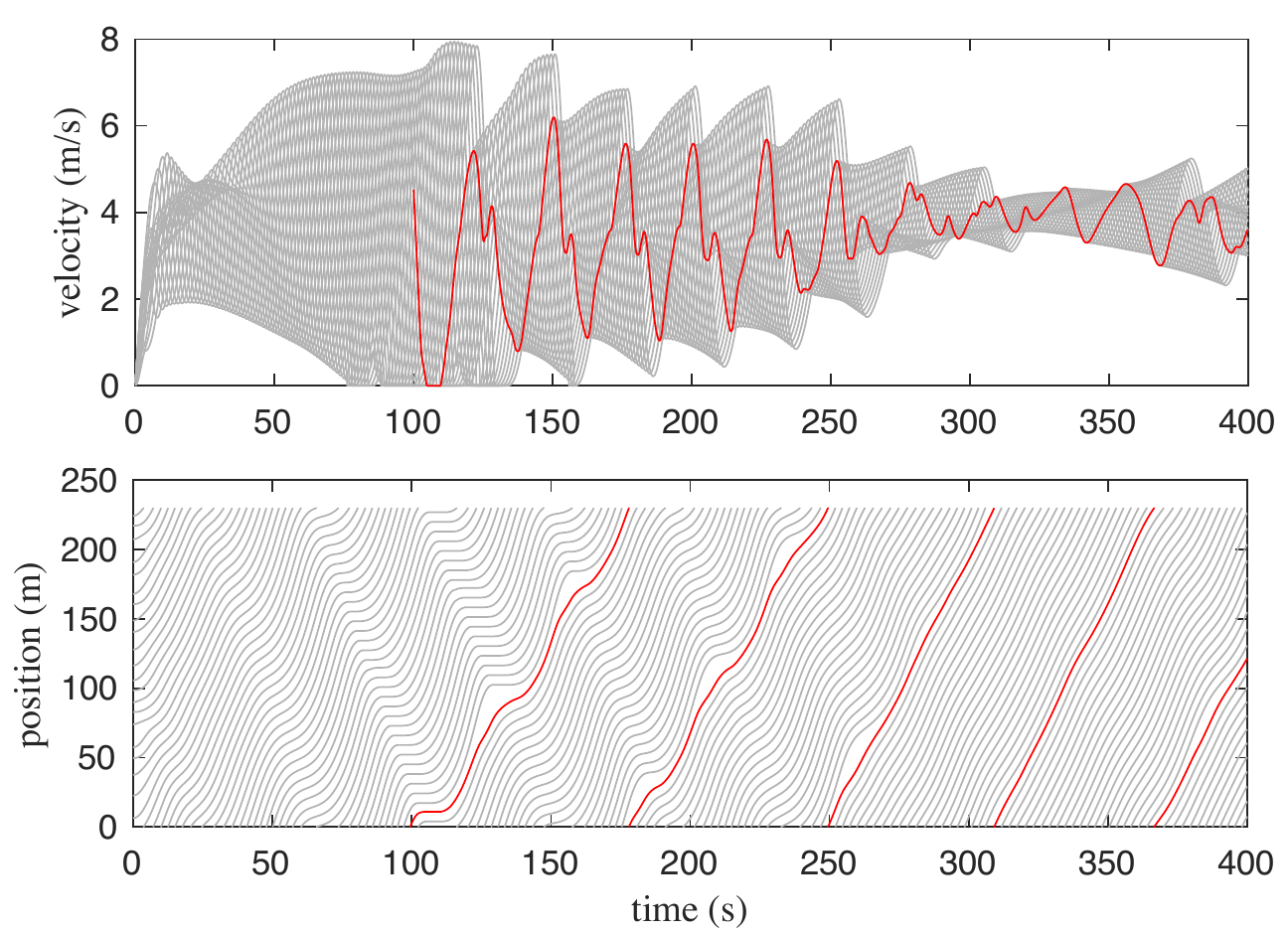}
    \caption{Top: speed profiles of all HDVs (grey) and the AV
(red) showing the performance of the backdoored AV controller. Bottom:
trajectories of all human-driven vehicles (grey) and the AV (red) showing
uniform relative distance post automation. The AV is controlled after 100
seconds as shown to be marked with red.}
    \label{ILL41}
\end{figure}
\begin{figure*}[h!]
    \centering
    \includegraphics[scale=0.65]{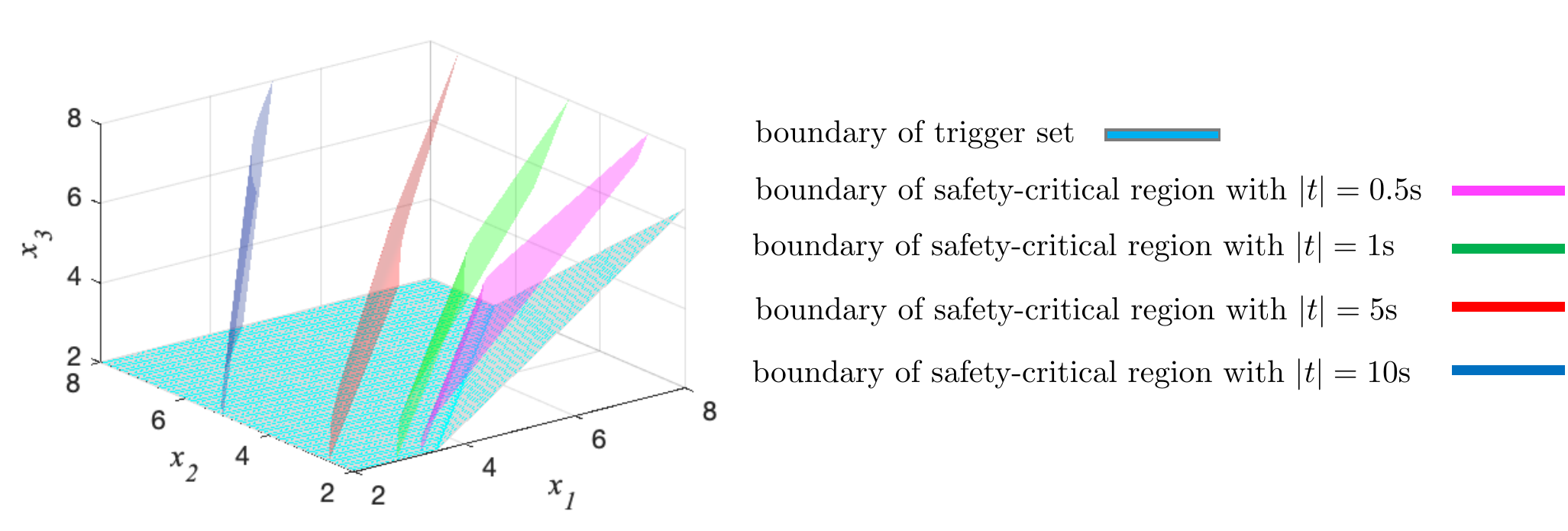}
    \caption{Security-critical areas (BRSs) with regards to different reach time $|t|$ in the original space. Notice that trigger set is security-critical area with $|t|=0$s. We only draw the boundary of set and the right hand side areas of each boundary are desired sets/regions.}
    \label{ILL5}
\end{figure*}
\begin{figure*}[h!]
    \centering
    \includegraphics[scale=0.65]{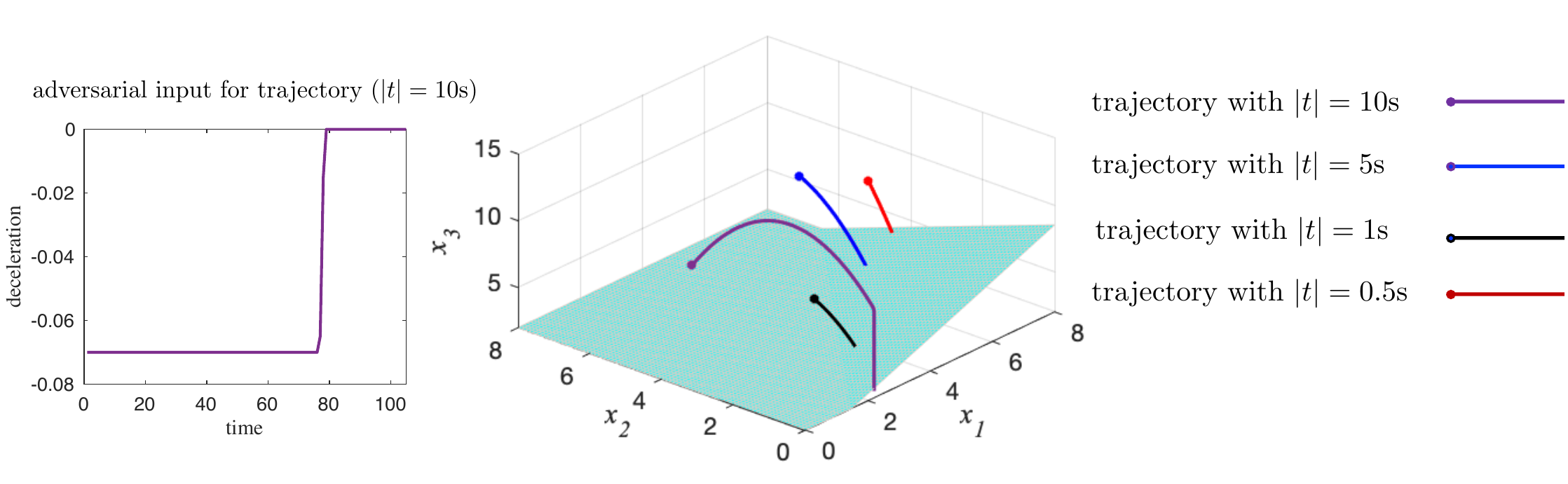}
    \caption{Left: adversarial control input for trajectory with $|t|=10$s, where the time step is $0.1$s; The adversarial input is $-0.07\mathrm{m}/\mathrm{s}^2$ and then close to $0\mathrm{m}/\mathrm{s}^2$; Right: Trajectories that start from different security-critical regions and enter trigger set.}
    \label{ILL6}
\end{figure*}
\begin{figure*}[h!]
    \centering
    \subfigure[]{
        \centering
        \includegraphics[scale=0.3]{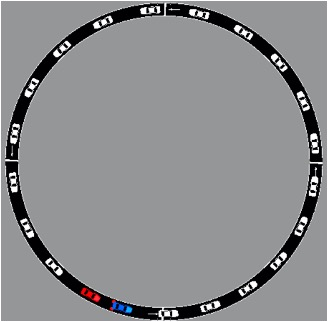}
        \label{ILL7a}}
    \subfigure[]{
        \centering
        \includegraphics[scale=0.3]{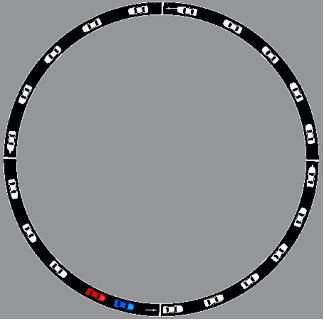}
         \label{ILL7b}} 
    \subfigure[]{
        \centering
        \includegraphics[scale=0.3]{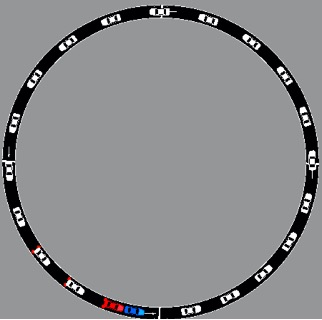}
         \label{ILL7c}}
    \subfigure[]{
        \centering
        \includegraphics[scale=0.68]{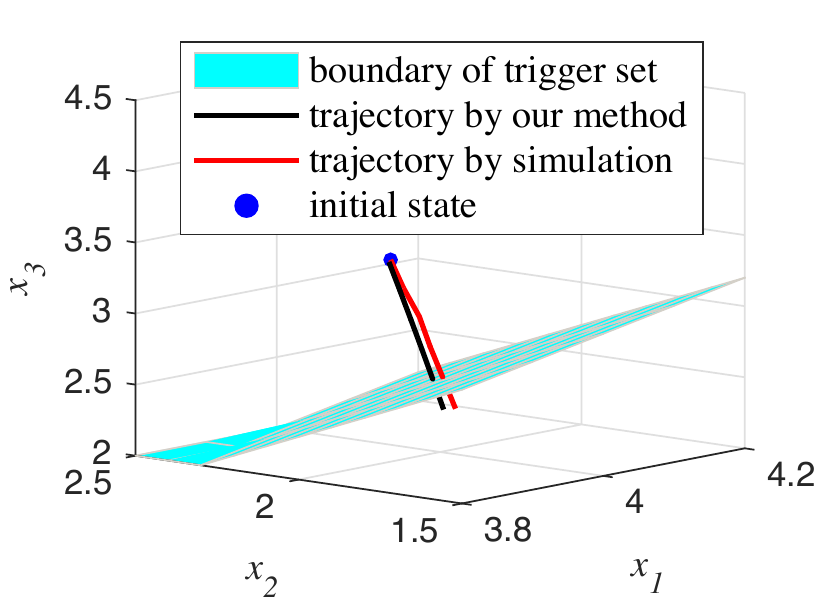}
         \label{ILL7d}} 
    \caption{Single-lane ring system with an AV (red) and an HDV (blue) manipulated by the adversary, where (a) shows the initial state; (b) shows the trigger states in 0.5s; (c) shows the collision state in 1s; (d) depicts the trajectories computed by our method and from simulation (which means feed the computed inputs into the DRL model of AV controller and interact with the traffic system).}
    \label{ILL7}
\end{figure*}
\section{Experimental Results}
\label{exp}
In this section we show the computational results. First, we demonstrate the (long-term) performance of Koopman operator approximation using a numerical example with known system dynamics. Then, the effectiveness of the proposed method for trigger activation is verified on a simulated traffic system  based on microscopic traffic simulator SUMO (Simulation
of Urban MObility) \cite{kheterpal2018flow}.

\subsection{A numerical example}
We consider the following linear discrete-time dynamic system:
\begin{equation}
    x_{k+1}=\mathbf{A}x_{k}, 
\end{equation}
with $k\in\mathbb{N}$ and a bounded $x_k$ (i.e., $\|x_k\|\leq a$), and 
\begin{equation}
    \mathbf{A}=\begin{bmatrix}

      1.0030 & 0.0000 & 0.0000 \\ 0.0008 & 0.9950 & 0.0998 \\ 0.0059 & -0.0998& 0.9950
    \end{bmatrix}
\end{equation}
 We do not consider the input since the long-term prediction performance majorly depends on $\mathbf{A}$ (as described in Section \ref{error}). We compute the LS solution by directly solving~\eqref{ls}, and obtain the stable solution by solving~\eqref{stab} using FGM \cite{gillis2019approximating} with default parameter settings. We run 10 simulations (each simulation 1000 data are sampled, where 500 data are used for training and 500 data for testing) and report the average results (with standard deviations). 

The eigenvalues of the stable matrix are $\{0.9970,0.9951,0.9930\}$. The results of propagated generalization errors  are presented in Fig.~\ref{pred}, where we notice that the  errors of stable solution do not increase dramatically compared with the LS solution when prediction step becomes large. We also present the one-step generalization errors of two solutions in Fig.~\ref{opre}, where we can see they have similar values. This implies a stable Koopman operator approximation can maintain a stable long-term performance.

\subsection{Results of reachability analysis based on SUMO}
\subsubsection{Traffic system based on SUMO}
Following \cite{wu2021flow}, our traffic systems are simulated
using the microscopic traffic simulator SUMO (Simulation
of Urban MObility), where 
all human-driven vehicles (HDVs) use the intelligent driver model
(IDM) as described in \cite{treiber2013traffic}. We use one AV for this system, where the controller of AV is trained by the deep deterministic policy gradient (DDPG) algorithm developed by \cite{lillicrap2015continuous}. The system indicated by eq.~\eqref{kpl} is discretized with a sampling time $\Delta t=0.1$s.
\subsubsection{DRL-based AV Controller}
We run our tests
on a circular system following the setting of Flow \cite{wu2017flow}, where 21 vehicles run on a 230 meters long a single lane, as presented in Fig.~\ref{ILL4}. We observe that traffic congestion as shown in Fig.~\ref{ILL4a} has been mitigated after adding one AV as presented in Fig.~\ref{ILL4b}, where vehicles become evenly spaced. There is also no collision after the addition of a AV, which is  illustrated by Fig.~\ref{ILL41} showing the speed profiles and space-time diagram of the system. This implies the DRL-based AV controller works well when suffering no backdoor attacks.

\subsubsection{Results of reachability analysis - backdoor trigger activation}
Following the work by \cite{wang2021stop}, a typical setting for attack interval ($t_{adv}$) and adversarial acceleration ($a_{adv}$) is $t_{adv}=1$s and $a_{adv}=0.5\mathrm{m}/\mathrm{s}^2$, so that $g(x)=x_2+x_3-x_1-0.5$ is the boundary of the trigger set in the original space. By choosing $d=3$, the lifted space is a $10$D space, and $g_{\Psi}(\Psi(x))=(x_2+x_3-x_1)^3-0.5^3$ is the boundary of trigger set in the lifted space. The results of both the backward reachable sets (BRSs) and trajectories are computed using the proposed method in the lifted space, and are plotted in the original $3$D space by recovering from the lifted space, unless otherwise stated.

Figure~\ref{ILL5} shows the trigger set and its corresponding BRSs (also known as security-critical areas) for different times $|t|$ in the original space. Only the boundaries of the BRSs are drawn, and the desired sets/regions are on the right-hand side of the boundaries.

After computing backward reachable sets (BRSs) for various reach times $|t|$, we randomly selected 1000 points from these BRSs as initial states. We then applied our control inputs to a deep reinforcement learning (DRL) model of an autonomous vehicle (AV) controller to determine if the initial states could be steered into the trigger set. The results are promising, as we were able to successfully trigger over $99\%$ of the randomly selected initial states within the computed BRSs. In Fig.~\ref{ILL6}, we depict the computed trajectories in the original space, which start from random initial states with regards to different time, i.e., $|t|=0.5,1,5,10$. We can observe that all trajectories enter the trigger set at the end, which implies the guiding information by our method is reliable to reach the trigger states (activate the triggers). We also present the computed optimal input from the adversary (actions of AV's leader) in the left subplot of Fig.~\ref{ILL6}, the adversarial input with regards to $|t|=10$s is to decelerate with $0.07\mathrm{m}/\mathrm{s}^2$ and then to maintain the speed until the trigger state is reached.

\begin{figure}[htbp!]
    \centering
        \includegraphics[scale=0.65]{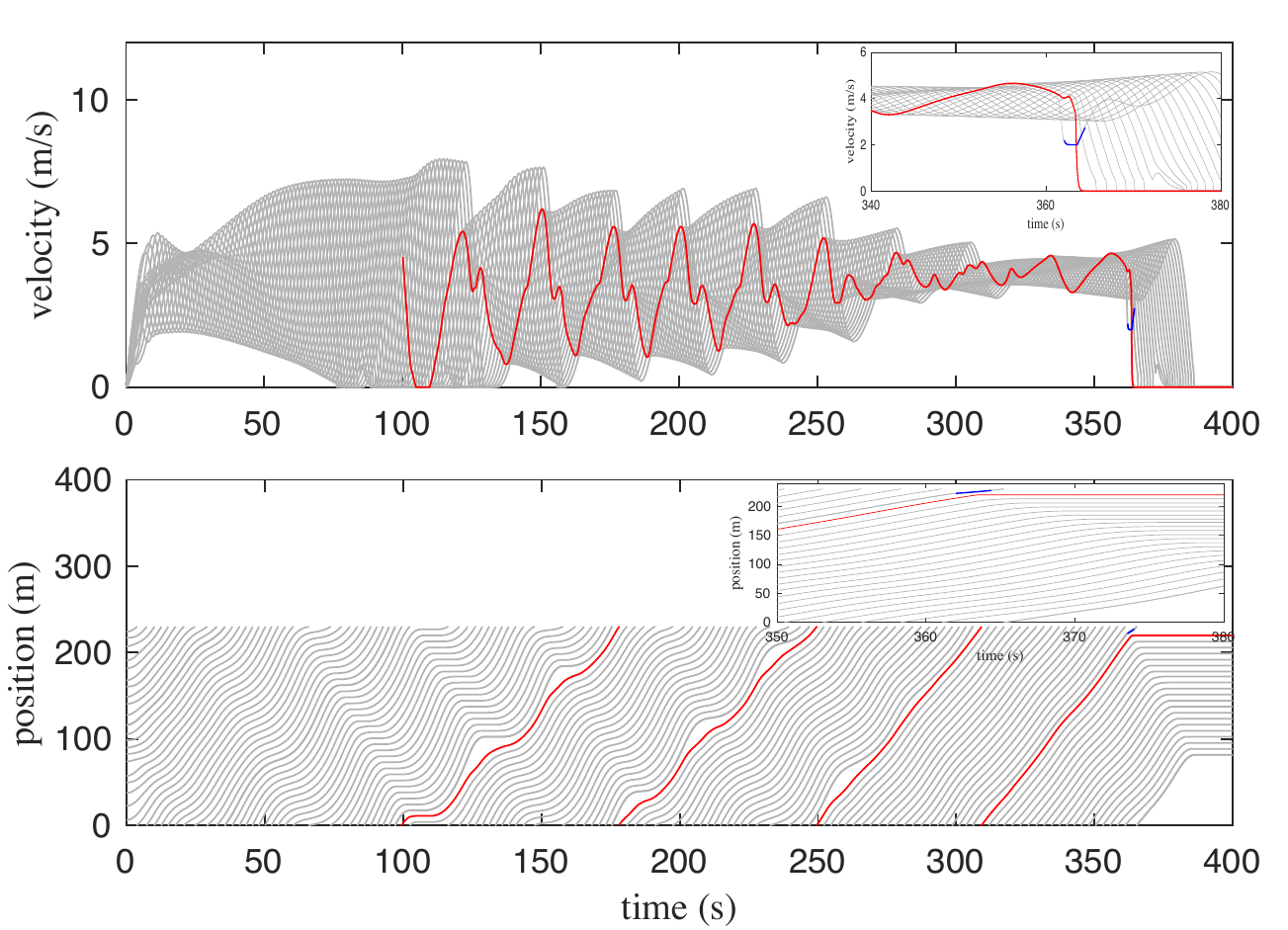}
    \caption{Top: speed profiles for backdoored model before and after attack; Bottom: spacing-time diagram before and after attack.The blue line in each figure denotes the trajectory that suffers adversarial inputs for a specific time period (i.e, 0.5s here). We observe that collision happens soon after the adversarial manipulation of HDV ahead of AV.}
    \label{ILL8}
\end{figure}

Fig.~\ref{ILL7} shows the simulation results of the single-lane ring system starting from a random initial state with achieving time $|t|=0.5$s. More specifically, Fig.~\ref{ILL7a} shows the initial state where all vehicles behave naturally with even spacing, Fig.~\ref{ILL7b} shows the trigger states 0.5s later, where the AV (red) has a larger speed (i.e., around 4$\mathrm{m}/\mathrm{s}$) than the HDV (blue, with speed around 2$\mathrm{m}/\mathrm{s}$) and their spacing is decreased (about  2.2 meters), and Fig.~\ref{ILL7c} shows collision happens around 1s later after the activation of trigger. Moreover, we present in Fig.~\ref{ILL7d} the associated trajectory computed by our method accompanied with the actual trajectory generated from simulation (using our guiding information), from which we observe these two trajectories stay close (especially the terminal states are close to each other), implying our guiding information is reliable.  For complementary, we draw the speed profile and spacing-time diagram in Fig.~\ref{ILL8} for the backdoored system before and after the trigger is activated. The adversarial input lasts for 0.5s (as indicated by the blue line) and then trigger is activated to cause a collision in around 1s (thereafter velocities of all vehicles decrease to 0 and positions of all vehicles do not change with time). This also implies our method is effective for trigger activation to cause collisions between the backdoored AV and its leading HDV.

\section{Conclusion and Future Work}
\label{con}
In this work, we introduce a novel data-driven approach to address the backdoor trigger activation problem. Specifically, we formulate the problem as a controlled dynamic system's reachability problem, and learn a surrogate linear dynamic model in the lifted space to efficiently compute the backward reachable set (BRS) while bounding the generalization error. Our approach identifies security-critical areas and adversarial inputs that can trigger the system into a compromised state, highlighting the vulnerability of the AV controller to backdoor attacks.
Furthermore, we demonstrate that our approach enables the design of efficient backdoor triggers that can cause vehicle collisions with the help of our guiding information. In our future work, we aim to focus on the defense side of the problem by constraining the objective of a deep reinforcement learning (DRL)-based controller to ensure that the traffic system operates within the unreachable sets of triggers. We believe that this will provide a defense mechanism against backdoor attacks and enhance the security of AV controllers.



\section*{Acknowledgment}
This work was supported by the NYUAD Center for Interacting Urban Networks (CITIES), funded by Tamkeen under the NYUAD Research Institute Award CG001.



%










\bibliographystyle{IEEEtran}
\bibliography{references}
\vspace{12pt}

\vfill


\end{document}